\documentclass[conference]{IEEEtran}
\IEEEoverridecommandlockouts
\usepackage{cite}
\usepackage{amsmath,amssymb,amsfonts}
\usepackage{algorithmic}
\usepackage{graphicx}
\usepackage{textcomp}
\usepackage{xcolor}

\def\BibTeX{{\rm B\kern-.05em{\sc i\kern-.025em b}\kern-.08em
    T\kern-.1667em\lower.7ex\hbox{E}\kern-.125emX}}
\begin{document}

\title{Content Based Player and Game Interaction Model for Game Recommendation in the Cold Start setting}

\author{\IEEEauthorblockN{Markus Viljanen, Tapio Pahikkala}
\IEEEauthorblockA{\textit{Department of Future Technologies} \\
\textit{University of Turku}\\
Turku, Finland
}
\and
\IEEEauthorblockN{Jukka Vahlo, Aki Koponen}
\IEEEauthorblockA{\textit{Centre for Collaborative Research} \\
\textit{University of Turku}\\
Turku, Finland
}
}

\maketitle

\begin{abstract}
Game recommendation is an important application of recommender systems. Recommendations are made possible by data sets of historical player and game interactions, and sometimes the data sets include features that describe games or players. Collaborative filtering has been found to be the most accurate predictor of past interactions. However, it can only be applied to predict new interactions for those games and players where a significant number of past interactions are present. In other words, predictions for completely new games and players is not possible. In this paper, we use a survey data set of game likes to present content based interaction models that generalize into new games, new players, and both new games and players simultaneously. We find that the models outperform collaborative filtering in these tasks, which makes them useful for real world game recommendation. The content models also provide interpretations of why certain games are liked by certain players for game analytics purposes.

\end{abstract}


\section{Introduction}

Game developers, publishers, and platforms are all interested to know why certain players play certain games and whether a game would be successful in the market place. Many game marketplaces also have the practical problem of recommending new games to players from a large catalog of possible games. Businesses are interested in these methods, because good recommendations can increase sales and user engagement. The academic field of recommender systems has investigated methods that can be trained on data sets of historical game and user interactions to predict new interactions. Recommender systems therefore seem like a natural solution to the game recommendation problem: gaming facilitates large data sets of past interactions and often additional information is available about game content and user profiles. 

Recommender systems have been investigated in different contexts, but the most popular algorithms can be divided into collaborative filtering (CF) and content based (CB) \cite{ref1}. Collaborative filtering is based on a data set of past player and game interactions, and it does not use player or game features because the predictions are made possible by observed correlations. For example, if players often play two games together, the games are probably similar and we can recommend one game when a player has played the other. In content based predictions, available information about games and/or players is used in enabling the predictions. In a game database, games typically have tags, genres, reviews, textual information, gameplay videos, etc. that can be used to create features. Creating player features is very flexible, because one can directly ask the players to answer questions about their preferences, motivations, and gaming habits. Predictions are then based on learning how game features or player features, or both together, result in the observed game likes. So called hybrid recommenders combine the content information (i.e. game features and player features) with collaborative filtering.

Academic research is often motivated by improving the accuracy of the methods, since this is an objective and easily evaluated task. However, the recommender system literature has started to recognize that accuracy may not always correlate with perceived utility  \cite{ref2}.
Collaborative filtering has been found to produce more accurate predictions, unless the methods are tasked to predict for players or games with few or no interactions \cite{ref3}. The setting with no interactions is known as the cold start problem, because collaborative filtering cannot predict in this setting. Research on game recommendation has evaluated methods by their predicted ability in historical player and game interactions, but in reality the problem of predicting for new games and new players is common. We therefore define four evaluation settings \cite{ref4}\cite{ref5}: predicting for past games and past players that have interactions (Setting 1), predicting for new games without players (Setting 2), predicting for new players without games (Setting 3), and predicting for new games and new players simultaneously (Setting 4).

In this study, we apply new methods to game recommendation that generalize better than collaborative filtering to the different settings. The resulting methods are simple, fast and easily interpretable. We therefore also interpret the model parameters, and find that they provide useful game market information about player traits such as gaming motivations and gameplay preferences. The development of the methods can be motivated by the standard approach to collaborative filtering, the Singular Value Decomposition (SVD), where one or both of the latent vectors is given. The task is then to learn the response of players or games, or their interactions, to the given features. The number of possible game and player pairs makes learning the interactions infeasible with the standard approach, so we utilize a mathematical result known as the vec-trick \cite{ref6} to train an identical model very fast.

\section{Related Work}

Our recommendation models produce a list of game recommendations to a player. This is known as Top-N recommendation, where the methods are evaluated on the accuracy of the predicted list of recommendations, in contrast to how well the methods would predict missing rating or playtime values for example. The objective of predicting an accurate list of game recommendations is probably the most relevant real world task of these systems, since this task has been adopted by many commercial companies. 

Top-N recommendation differs from the traditional task of predicting missing values. The evaluation is based on ranking accuracy metrics such as precision or recall, not on error metrics like the root mean squared error (RMSE). There is no guarantee that algorithms optimized for the RMSE are also optimal for ranking because the task is different. Furthermore, item popularity has been found to have a large effect on the error metrics \cite{ref8}. The k Nearest Neighbour (kNN) and the Singular Value Decomposition (SVD) based matrix factorization have become standard methods in predicting missing values, but they also work for the top-N recommendation task \cite{ref3}. 

In the task of producing game recommendations, one should also take into account whether information about the player-game interactions is implicit or explicit. Implicit signals about liking a game are for example owned or played games, whereas explicit data is the ratings and opinions provided about the games. Implicit data is often complete, which means that every player and game pair has a value and the task is to rank the games by the probability of liking them. In explicit data, there are typically many missing values because a player has not rated every possible game, and the task is to predict the values of the missing ratings. 

For our models we technically use explicit data, because the player and game interactions are based on the favourite games mentioned in a survey. Similar data sets to our explicit survey based data on favorite games and player preferences could be obtained without using surveys, i.e. implicitly by crawling gaming platforms for the games that users own or play. In either case, these data sets are understood as complete data for our task. This means that there are no missing values because all player and game pairs have a value that denotes whether the player mentioned, owned or played the game.

In this study, also we utilize player traits and preferences to construct player features. In game research, player preferences can be divided into four main categories: 1) player motivations \cite{x1, x2}, 2) preferred play styles \cite{x3, x4}, 3) gaming mentalities \cite{x5, x6}, and 4) gameplay type preferences \cite{x7, x8}. Player motivations measure general reasons why players play games, whereas models that investigate play styles are typically based on player behavior data rather than survey data. Gaming mentalities refer to the psychometric data on players' typical and preferred gaming intensity type (e.g. casual or hardcore gaming). \cite{x9} Of the four approaches on player preferences, gameplay type preference data is arguably the most promising for producing personalized game recommendations, because it is closely related to game features. Furthermore, it has been been shown that gameplay type preferences such as preference in exploration, management or aggression predicts habit to play games of specific genres  \cite{x7}. Because of these reasons, we make use of gameplay type preference survey data in this study.

Comparing recommender systems is difficult for several reasons. The performance may depend on the data set, the prediction task, and the chosen metric \cite{ref2}. In addition, many methods are sensitive to the choice of hyperparameters and the optimization method, which means that authors of new methods may have not always used the best baselines \cite{ref10}. There can also be  performance differences between different software implementations of otherwise identical methods \cite{ref11}. However, the simple baseline methods tend to produce competitive results when the hyperparameters are carefully tuned \cite{ref10}. High accuracy if often assumed to correlate with an useful recommender system, but subjective utility recommender system has also become an important research goal in itself \cite{ref2}. Optimizing accuracy can lead to recommending popular items at the cost of personalized results \cite{ref8}. 

There are studies that have investigated the development of new recommender systems to games. The first study \cite{ref12} used probabilistic matrix factorization based collaborative filtering for the Xbox platform. The second study \cite{ref13} presented a recommender based on archetypes, where their formula (5) is a constrained case of the SVD. The study included comparisons to kNN (cosine) trained on the latent SVD factors, the standard kNN with somewhat small neighbourhood sizes, and random or most popular recommendations. The third study \cite{ref14} investigated a new case-based disability rehabilitation recommender, which is a type of content recommender. The content was based on game descriptions combined with social network information and questionnaire answers of the users. The fourth study \cite{ref15} presented a graph based approach with a biased random walk model inspired by ItemRank, which is a type of a hybrid recommender. The fifth study \cite{ref16} is a kNN (cosine)  recommender based on content created by the latent semantic analysis of wikipedia article. The sixth study \cite{ref17} defined a content based recommender to find similar games through the cosine similarity of feature vectors based on Gamespot game reviews. The bag-of words representation was outperformed by information theoretic co-clustering of adjective-context word pairs to reduce the dimensionality. The seventh study evaluated the quantitative and qualitative performance of game recommenders on the Steam data set  \cite{ref18}. They used BPR++, grouped BPR, kNN (cosine) on game tag content, kNN combined with grouped BPR, a simulation of Steam's recommender, SVD, Factorization Machine (FM), and popularity ranking. They found significant quantitative differences but no qualitative differences between the methods.

\section{Data Set}

\subsection{Survey data set}

The survey data (N=15,894) was collected by using a total of 10 standalone web-based surveys. Each of the surveys focused on different aspects of player preferences but all surveys included open-ended questions about respondents favorite games. Most of the samples were collected by using a UK-based crowd sourcing platform, market companies providing large online panels, or by recruiting respondents from social media platforms such as Facebook or Reddit. The survey data includes representative samples from Finland, Denmark, USA, Canada, and Japan. A typical survey took up to 20 minutes to complete with a computer or a mobile phone, and was targeted to everyone between the ages of 18 and 60. Another type of survey data was collected by using a short online player type test. The short test was open to everyone regardless of their prior experience in playing games, or the possible lack of it. Before analyzing survey data of the individual survey data sets, the data was cleaned of participants who implied content non-responsivity by responding similarly to every question. 

The survey data analyzed in this study consists of 15894 observations, 6465 unique games, and 80916 favorite game mentions. There are 1 to 37 favorite game mentions per player, and on average a player mentioned 5 games as his or her favorites. Every game that is mentioned as a favorite game has from 1 to 1108 individual favorite game mentions i.e. players. The data set is very sparse, since only 0.08\% of possible (player, game)-pairs are favorite game mentions. The answers tended to be more novel and personal than data sets based on played or owned games.

Content for the games was obtained by crawling game tags from publicly available sources, such as the Steam platform and the Internet Games DataBase (IGDB), the latter of which provided an API for this purpose. The presence or absence of every tag in each game was stored as a binary indicator. 

Content for the players was obtained by asking the survey participants' preferences using the Gameplay Activity Inventory (GAIN) which has been validated with cross-cultural data. The GAIN consists of 52 questions (5-Likert scale, 1=very unpleasant, 5=very pleasant) about gameplay preferences, and the inventory measures five dimensions of gameplay appreciation: aggression, management, caretaking, exploration, and coordination. These questions consists of items such as 'Exploding and Destroying' (factor: aggression), 'Searching for and collecting rare treasures' (factor: exploration), 'Flirting, seducing and romantic dating' (factor: caretaking), 'Matching tiles together' (factor: coordination), 'Generating resources such as energy or money' (factor: management) etc. (Vahlo et al. 2018). In addition to the GAIN model, we utilized also a 9-item inventory on players' game challenge preferences. These items measure how pleasurable players consider e.g. 'logical problem-solving', 'strategic thinking', 'puzzle solving', 'racing against time', etc. A typical survey included the full 52 plus 9 questions whereas the online player type test consisted of a partly randomized sample of these questions. 



\section{Models}

\subsection{Data set and validation}

Assume we have $n$ players and $m$ games. Denote player $i\in\{1,2,...,n\}$ and game $j\in\{1,2,...,m\}$. The player and game interactions are stored in a $n \times m$ binary game like matrix $R_{i,j}=\mathbb{I}(\text{player }i\text{ likes game }j)$. The matrix does not have missing entries, because the game like status is known for every player. For example, the player $i$ may have answered the first and the third game as their favourites:

\begin{equation}
\begin{array}{c}
R_{i,:} = (1, 0, 1, 0, ..., 0)
\end{array}
\end{equation}

The task is to predict the ranking of games that the player has not answered as their favourite but might like. These predictions are the matrix $R_{i,j}^{*}\in\mathbb{R}$, where only the order of the values in each row matters for ranking. For example, the predictions for player $i$ over all the $m$ games could be:

\begin{equation}
\begin{array}{c}
R_{i,:}^{*} = (1.41, 0.10, 0.82, 0.04, ..., 0.21)
\end{array}
\end{equation}

The recommendation list for a player is obtained by taking the indices of games with $k$ largest predicted values in $R_{i,:}^{*}$, where the games that the player has already liked are excluded.

In addition to game likes, we also have player and game features  that we can use for content based prediction. Denote the $m \times r$ matrix of game features as $X_{\text{tags}}$, where the feature vectors for the $m$ games are stored as rows. In our case these features are indicators of presence or absence of each of the $r$ game tags. Denote the $n \times s$ matrix of player features as $X_{\text{questions}}$, where the feature vectors for the $n$ players are stored as rows. In our case these features are the responses to the $s$ questions on a Likert scale of -2,-1,0,1,2. 

\begin{figure}[htbp]
\centerline{\includegraphics[width=\columnwidth]{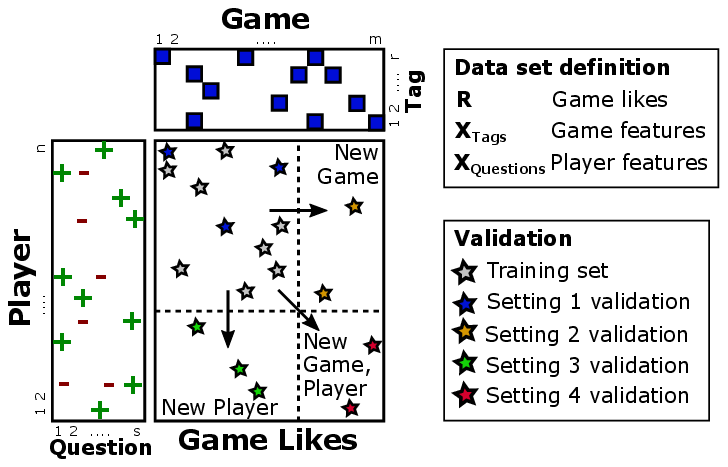}}
\caption{Illustration of the data set and validation settings.}
\label{figure:data}
\end{figure}

To test model performance, we split the data set into training and validation sets as follows. First, we randomly sampled 25\% of games into 'test games' an 25\% of players into 'test players' that the models do not see during training. These games and players test the performance of the model for new games and players. The other games and players belong to the training set. In Setting 1, the models are tested on the task of recommending known games for a known player who has mentioned 3 favourite games. We therefore further selected 20 \% of the training set players by randomly sampling amongst those who have liked more than 3 games. Randomly chosen 3 games of each player are the 'seed' which belongs to the training set, and the remaining games for these players belong to the validation set for Setting 1. The Setting 2 (new games) validation set consists of the unseen 'test games' for the known players. In Setting 3 (new players), the validations set consists of unseen 'test players' for the known games. In setting 4, the validation set is the game likes for the unseen 'test games' and 'test players'. We have illustrated the game like matrix $R$, the game features $X_{\text{tags}}$, the player features $X_{\text{questions}}$, and the different validation settings in Figure~\ref{figure:data}.

\subsection{Metrics}

We use Precision@20 and nDCG@m to measure accuracy in the validation sets. They are defined as follows.

\subsubsection{Precision@20}

The Precision@k metric counts the number of games the player has liked in the validation set, as a fraction of all games in a recommendation list of length $k$. Assume the model predicts $R_{i,:}^{*}$ for player $i$, and denote $r^{(i)}$ as the vector of indices that sorts the predictions. The element $r^{(i)}_{j}$ is the $j$'th game in the recommendation list, i.e. the index of $j$'th largest predicted value in $R_{i,:}^{*}$. The metric for the data set is the average precision over the players:

\begin{equation}
\begin{array}{c}
\frac{1}{n}\sum_{i=1}^{n}\frac{1}{k}\sum_{j=1}^{k} \mathbb{I}(\text{player } i \text{ likes game } r^{(i)}_{j})
\end{array}
\end{equation}

Precision is a realistic measure of the real-world accuracy of a recommendation list, where $k$ is typically small and the position of a game in the list does not matter.

\subsubsection{nDCG@k}
The normalized Discounted Cumulative Gain (nDCG@k) metric measures the position of liked games in the recommendation list. When a player liked a game, its position in the player's recommendation list is rewarded by the inverse of its logarithmic rank. These are called the discounted cumulative gains. In the optimal ranking, we have ranked liked games on the top of the recommendation list, of the total $k_i=|\{j : R_{i,j}^{\text{validation}} = 1\}|$ liked games, and the discounted cumulative gain has the value $\text{IDCG}_i=\sum_{j=1}^{\text{min}(k_i,k)}1/\text{log}_2(j+1)$. The nDCG@k is the discounted cumulative gain in the recommendation list $r^{(i)}$ of length $k$, normalized by the maximum attainable value $\text{IDCG}_i$. The metric for the data set is the average over the players:

\begin{equation}
\begin{array}{c}
\frac{1}{n}\sum_{i=1}^{n}\frac{1}{\text{IDCG}_i}\sum_{j=1}^{k} \frac{\mathbb{I}(\text{player } i \text{ likes game } r^{(i)}_{j})}{\text{log}_2(j+1)}
\end{array}
\end{equation}

Because Precision@20 measures the top recommendations, we used used nDCG@m to measure the overall ranking.

%


\subsection{Models}

\subsubsection{Multivariate Normal Distribution (MVN)}

First we present a simple new collaborative filtering model which has a competitive accuracy but simpler interpretation. This model allows us to explain the recommendations through explicit correlation matrix between games. The model also allows us to completely remove the influence of game popularity to investigate its effect. Assume that every row of the player like matrix is a sample from a multivariate normal distribution: $R_{i,:} \sim\mathcal{N}(\mu,\Sigma)$ with mean vector $\mu\in\mathbb{R}^{m}$ and covariance matrix $\Sigma\in\mathbb{R}^{m \times m}$. This model has a closed form solution for the distribution parameters, because the maximum likelihood estimate of these is the sample mean vector $\mu_j = \frac{1}{n}\sum_{i} R_{i,j}$ and the sample covariance matrix $\Sigma_{i,j} = \frac{1}{n}\sum_{s}(R_{s,i}-\mu_i)(R_{s,j}-\mu_j)$.

In prediction time, we assume that the values of liked games are known to be one but the values for other games are missing. Denote the indices of liked games as $\mathcal{I}$ and the indices of other games as $\mathcal{J}$ so that $\mathcal{I}\cup\mathcal{J}=\{1,2,...m\}$. We use indexing $X_{\mathcal{J},\mathcal{I}}$ to denote the submatrix with rows from $\mathcal{J}$ and columns from $\mathcal{I}$, for example. The predictions for the missing game likes are then given by the expectation of the conditional distribution $R_{i,\mathcal{J}}^{*} = \mathrm{E}(R_{i,\mathcal{J}}|(R_{i,j}=1)_{j\in\mathcal{I}})$. This can be shown to equal the solution: 

\begin{equation}
\begin{array}{c}
R_{i,\mathcal{J}}^{*} = \mu_{\mathcal{J}} 
+ \Sigma_{\mathcal{J},\mathcal{I}}(\Sigma_{\mathcal{I},\mathcal{I}})^{-1}(\overline{1}-\mu_{\mathcal{I}})
\end{array}
\end{equation}

To predict without game popularity, we remove the mean vectors from the formula and substitute the sample covariance matrix with the sample correlation matrix. This is equivalent mean centering and then normalizing the game like matrix column wise before applying the model.

\subsubsection{k Nearest Neighbour (kNN)}

The kNN is a simple recommendation method. Assume we have calculated the similarity between any two games in a $m \times m$ matrix $S_{ij}$. The rating prediction for game $j$ considers the $k$ most similar games in the game like matrix for player $i$. Denote this set of most similar games $D^{k}(i,j)$. The prediction for a player is the similarity weighted average to the player's like status of $k$ most similar games: $R_{i,j}^{*} = \sum_{s\in D^{k}(i,j)} S_{j,s} R_{i,s}/\sum_{s\in D^{k}(i,j)} S_{j,s}$. We evaluated the kNN on neigbhourhood sizes of $k=1,2,4,8,...,m$, but we always obtained the best results with the maximum neighbourhood size of $k=m$. As others have pointed out \cite{ref8}, the normalizing denominator is not necessary for the ranking task and we in fact obtained better predictions without it. We therefore predict simply by:

\begin{equation}
\label{similarity}
\begin{array}{c}
R_{i,j}^{*} = \sum_{s\in D^{k}(i,j)} S_{j,s} R_{i,s}
\end{array}
\end{equation}

We define the similarity function as either the cosine (cos) or the Pearson correlation (phi), where $\mu_j=\sum_{i} R_{i,j}/n$ is the column mean of the game like matrix:

\begin{equation}
\begin{array}{c}
S_{i,j}^{\text{cos}}=\frac{\sum_{s}R_{s,i}R_{s,j}}{\sqrt{\sum_{s}R_{s,i}^2}\sqrt{\sum_{s}R_{s,j}^2}}
\end{array}
\end{equation}

\begin{equation}
\begin{array}{c}
S_{i,j}^{\text{phi}}=\frac{\sum_{s}(R_{s,i}-\mu_i)(R_{s,j}-\mu_j)}{\sqrt{\sum_{s}(R_{s,i}-\mu_i)^2}\sqrt{\sum_{s}(R_{s,j}-\mu_j)^2}}
\end{array}
\end{equation}

\begin{figure*}
  \includegraphics[width=\textwidth]{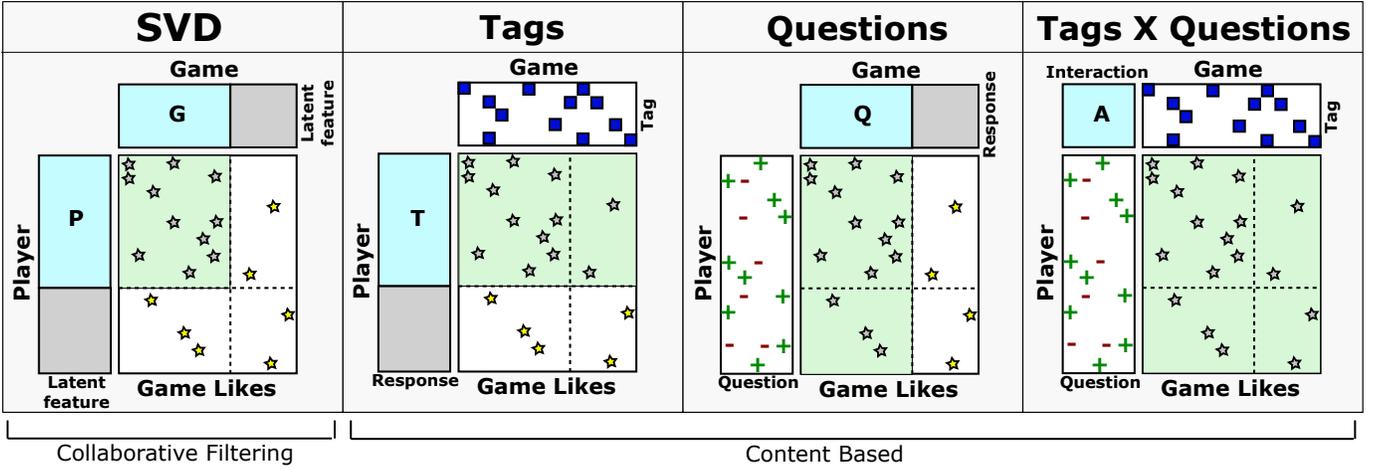}
  \caption{Illustration of the SVD and content models. Models can only generalize to settings (light green) for which they learn representations (light blue).}
  \label{fig:models}
\end{figure*}

\subsubsection{Singular Value Decomposition (SVD)}

The SVD of dimension $k$ is defined in terms of $n \times k$ matrix $P$ of latent player factors as rows and $m \times k$ matrix $G$ of latent game factors as rows. It is a type of regularized matrix factorization because the predicted game like matrix is the product $R^{*}=PG^T$ of the factor matrices. A prediction for player $i$ and game $j$ is simply the product of the latent user vector $P_{i,:}\in\mathbb{R}^k$ and the latent game vector $G_{j,:}\in\mathbb{R}^k$. These latent vectors are initially unknown. We evaluated different choices of dimension $k=4,8,16,32,64,128$ and regularization parameter $\lambda$. For the SVD implementation in the Suprise library, a python package for implicit recommendations, we found that a grid of $\lambda=1,2,4,8,16,32,64,128,256,512$ produced a concave maximum between the values. We call the choice of $\lambda=0$ as PureSVD, because it is possible to use a standard singular value decomposition. The SVD was sensitive to regularization, but if the regularization choice was optimal we obtained almost identical results for dimensions $k\geq32$. The model parameters are found by minimizing the RMSE between observed game likes and predicted game likes: 

\begin{equation}
\begin{array}{c}
P, G = \text{argmin}_{P,G} \|R-PG^T\|^2_F + \lambda \|P\|^2_F + \lambda \|G\|^2_F
\end{array}
\end{equation}

Where the matrix norm $\|X\|^2_F$ denotes the Frobenius norm, or the root mean square of every element in the matrix $X$. To find the parameters, one approach is to use Alternating Least Squares (ALS) optimization \cite{ref3}. In this method, either the latent game vectors $G$ or the latent player vectors $P$ are assumed to be fixed and the optimal solution for the other is found. Because in this case every row independently minimizes the squared error associated with that row, we can solve for each row with standard ridge regression either:

\begin{equation}
\begin{array}{c}
P_{i,:} = (G^T G + \lambda I)^{-1} G^T R_{i,:}^T
\end{array}
\end{equation}

\begin{equation}
\begin{array}{c}
G_{j,:} = (P^T P + \lambda I)^{-1} P^T R_{:,j}
\end{array}
\end{equation}

The optimization starts by initializing $P$ and $G$ with random values. We iterate between fixing $G$ to find optimal values for $P$, and then fix the resulting $P$ to find optimal values for $G$. This is repeated until convergence.

\subsubsection{Tags}

The first content model is based on game features, which we call the 'Tags' model because our game features are based on game tags. We  assume that each player has some interaction strength with each game feature. These interaction strengths are described by a player specific vector of length $r$. Collect these vectors as rows of the $n \times r$ model parameter matrix $T$, which needs to be learned from data. A given player may for example answer that they like 'Candy Crush' and 'Tetris', which implies that the player interacts positively with game tags 'puzzle' and 'tile-matching'. We predict the game likes as a product of the game features $X_{\text{tags}}$ and the player interaction strengths $T$: $R^{*}=TX_{\text{tags}}^T$. To find the parameters, we minimize the RMSE between observed game likes and predicted game likes:

\begin{equation}
\begin{array}{c}
T = \text{argmin}_{T} \|R- TX_{\text{tags}}^T\|^2_F + \lambda \|T\|^2_F
\end{array}
\end{equation}

Every row $T_{i,:}$ in fact independently minimizes the squared error associated with that row, so the model can be fitted separately for every row with standard ridge regression:

\begin{equation}
\begin{array}{c}
T_{i,:} = (X_{\text{tags}}^T X_{\text{tags}} + \lambda I)^{-1} X_{\text{tags}}^T R_{i,:}^T
\end{array}
\end{equation}

\subsubsection{Questions}

The second content model is based on player features, which we call the 'Questions' model because our player features are based on questionnaire about gaming preferences. We assume that each game has an interaction strength with each player feature. These interaction strengths are described by a game specific vector of length $s$. Collect these vectors as rows of the $n \times s$ model parameter matrix $Q$, which needs to be learned from data. A given game 'Candy Crush' may for example be liked by players that have stated a preference for 'logical challenges' and 'racing against time'.  We predict the game likes as a product of the player features $X_{\text{questions}}$ and the game interaction strengths $Q$: $R^{*}=X_{\text{questions}} Q^T$. To find the parameters, we minimize the RMSE between observed game likes and predicted game likes:

\begin{equation}
\begin{array}{c}
Q = \text{argmin}_{Q} \|R- X_{\text{questions}} Q^T\|^2_F + \lambda \|Q\|^2_F
\end{array}
\end{equation}

Again, every row $Q_{i,:}$ independently minimizes the squared error associated with that row, so the model can be fitted separately for every row of $Q$ with standard ridge regression:

\begin{equation}
\begin{array}{c}
Q_{j,:} = (X_{\text{questions}}^T X_{\text{questions}} + \lambda I)^{-1} X_{\text{questions}}^T R_{:,j}
\end{array}
\end{equation}

\subsubsection{Tags X Questions}

The final content model is based on both game and player features, which we call the 'Tags x Questions' model because it is based on interactions between the game tags and the questionnaire answers. To model the game likes between every player $i$ and every game $j$, we assume that each (player, game)-pair is described by a feature vector. The feature vector for the pair is the tensor product of the player's features and the game's features. Given the $n \times s$ player feature matrix $X_{\text{questions}}$ and the $m \times r$ game feature matrix $X_{\text{tags}}$, the pair feature matrix can be represented as a $nm \times sr$ feature matrix $X_{\text{questions}} \otimes X_{\text{tags}}$. The logic behind this representation implies that the game likes are simply the sum of interaction strengths between player and game features, where the $r \times s$ interaction strength matrix as $A$ needs to be learned from data. A given player may for example answer that they like 'logical challenges' and 'racing against time', and the data implies that these answers interact positively with game tags 'puzzle' and 'tile-matching'. Denote the columnwise stacking of the interaction strength matrix as $\text{vec}(A)$, and the columnwise stacking of the $n \times m$ game like matrix as $\text{vec}(R^{*})$. Further denote the feature matrix as $X_{\text{interactions}}=X_{\text{questions}} \otimes X_{\text{tags}}$. The response is predicted as the sum of the player feature and game feature interactions: $\text{vec}(R^{*})=X_{\text{interactions}}\text{vec}(A)$. To find the model parameters, we minimize the RMSE between observed game likes and predicted game likes:

\begin{equation}
\begin{array}{c}
A = \text{argmin}_{A} \|\text{vec}(R^{*})-X_{\text{interactions}} \text{vec}(A)\|^2_F + \lambda \|A\|^2_F
\end{array}
\end{equation}

There is a mathematical optimization shortcut known as the vec-trick \cite{ref6}, which makes the minimization problem computationally tractable in the special case that the feature matrix is a Kronecker product. 
We use the python software package RLScore\footnote{http://staff.cs.utu.fi/~aatapa/software/RLScore} which implements this short cut to obtain an exact closed form solution to the ridge regression problem:

\begin{equation}
\begin{array}{c}
\text{vec}(A) = (X_{\text{interactions}}^T X_{\text{interactions}} + \lambda I)^{-1} X_{\text{interactions}}^T \text{vec}(R^{*})
\end{array}
\end{equation}

\section{Results}

\subsection{Model accuracy}

\begin{table}[htbp]
\caption{Ranking Accuracy by nDCG@m (\%)}
\label{nDCG}
\begin{center}
\begin{tabular}{|r|r|r|r|r|}
\hline
\textbf{Model} & \textbf{Setting 1} & \textbf{Setting 2} & \textbf{Setting 3} & \textbf{Setting 4} \\
\hline
Random & 13.9 & 13.2 & 14.6 & 13.4 \\
\hline
MVN & \textbf{31.9} & 13.2 & 14.6 & 13.4 \\
\hline
kNN (cos) & 30.0 & 13.2 & 14.6 & 13.4 \\
\hline
kNN (phi) & 27.4 & 13.2 & 14.6 & 13.4 \\
\hline
PureSVD & 28.4 & 13.2 & 14.6 & 13.4 \\
\hline
SVD & 30.9 & 13.2 & 14.6 & 13.4 \\
\hline
Tags & 23.4 & \textbf{22.3} & 14.6 & 13.4 \\
\hline
Questions & 26.9 & 13.2 & \textbf{32.2} & 13.4 \\
\hline
Tags X Questions & 23.2 & 19.9 & 24.3 & \textbf{20.1} \\
\hline
\end{tabular}
\label{tab1}
\end{center}
\end{table}

\begin{table}[htbp]
\caption{Recommendation List Accuracy by Precision@20 (\%)}
\label{Precision}
\begin{center}
\begin{tabular}{|r|r|r|r|r|}
\hline
\textbf{Model} & \textbf{Setting 1} & \textbf{Setting 2} & \textbf{Setting 3} & \textbf{Setting 4} \\
\hline
Random & 0.1 & 0.1 & 0.1 & 0.1 \\
\hline
MVN & \textbf{5.0} & 0.1 & 0.1 & 0.1 \\
\hline
kNN (cos) & 3.9 & 0.1 & 0.1 & 0.1 \\
\hline
kNN (phi) & 3.2 & 0.1 & 0.1 & 0.1 \\
\hline
PureSVD & 3.8 & 0.1 & 0.1 & 0.1 \\
\hline
SVD & 4.3 & 0.1 & 0.1 & 0.1 \\
\hline
Tags & 2.3 & \textbf{1.7} & 0.1 & 0.1 \\
\hline
Questions & 3.4 & 0.1 & \textbf{4.3} & 0.1 \\
\hline
Tags X Questions & 2.4 & 1.1 & 2.7 & \textbf{1.2} \\
\hline
\end{tabular}
\label{tab1}
\end{center}
\end{table}

We report the accuracy of different models using nDCG@m in Table~\ref{nDCG} and Precision@20 in Table~\ref{Precision}. The metrics have values between 0-100\%, yet it is challenging to interpret the quality of recommendations from their absolute values. We can use accuracy metrics to compare and improve the models, but the results need to be verified qualitatively in practise. 

Comparing models by accuracy tells a clear-cut story. The two metrics have the same interpretation. In Setting 1, collaborative filtering methods outperform content based approaches with the MVN model delivering the best results. In Setting 2 we generalize to the new games, and only the Tags and Tags X Questions models are able to generalize. The Tags model has a greater degree of freedom and it performs better in this setting. In Setting 3 we generalize to new players, and only the Questions and Tags x Questions models are able to generalize. Again, the Questions model has more freedom to fit the data and performs better in this setting. For the last setting, only the Tags X Questions model which uses both features is able to make useful predictions. 

The mathematics underlying the generalization ability is illustrated in Figure~\ref{fig:models}. Predictions can only be made for player and game pairs where both games and players have representations which have been learned from the training set. However, if the representations can be learned for the setting it is useful to use more flexible models. There is therefore an important trade-off between using the provided features to generalize better or learning latent features to have better performance inside the training set games and players.

\subsection{Model interpretation}

\begin{figure*}
  \includegraphics[width=\textwidth]{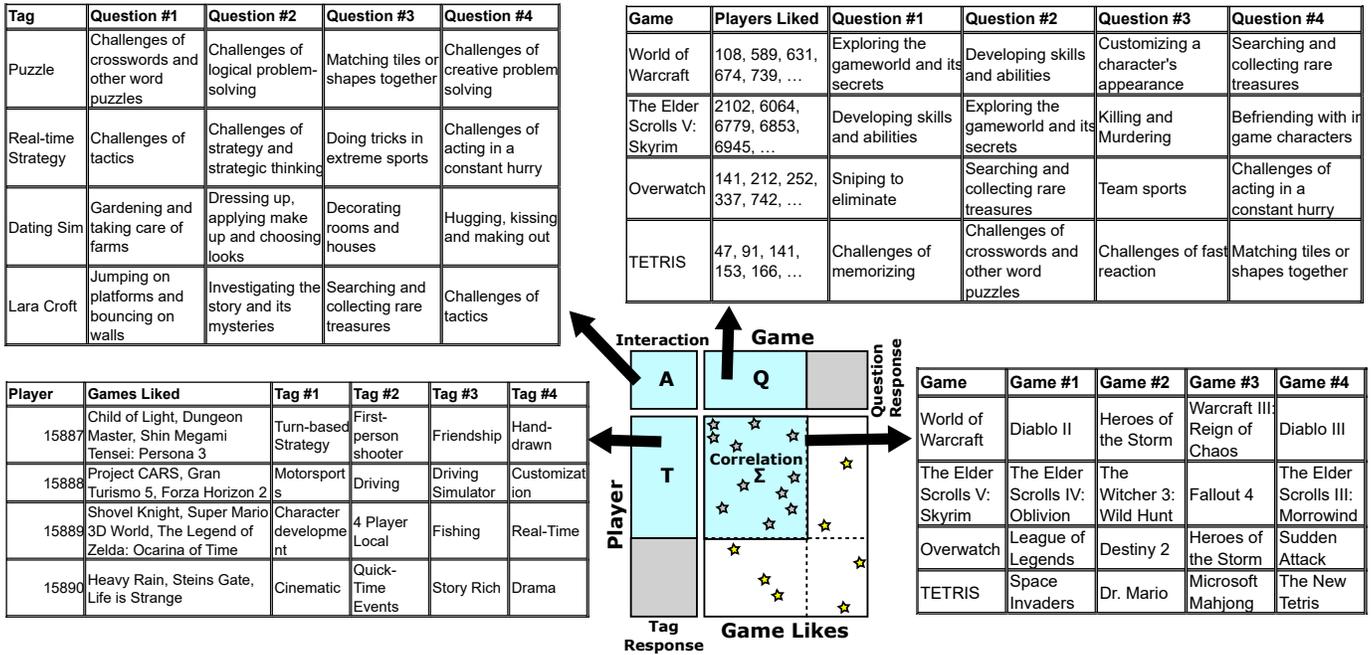}
  \caption{Example of model interpretation: correlated games (MVN), tag responses (Tags), question responses (Questions) and interactions (Tags x Questions)}
  \label{fig:interpretation}
\end{figure*}

Because the models are based on simple linear models, we can interpret the model coefficients. The coefficients provide an explanation of why certain players are predicted to like certain games. This information can be useful for game developers or publishers that seek to understand the gaming market, and they can be used to tune the model towards qualitatively better predictions.

Recalling that the MVN model predictions are based on the game mean vector $\mu$ and covariance matrix $\Sigma$, we interpret how collaborative filtering arrives at the predictions. The mean vector is simply the popularity of each game, calculated as the fraction of players that play each game. This is the starting point of the predictions, which are then adjusted based on the strength or positive or negative correlations to the games the player has played. For example, in Figure~\ref{fig:interpretation}, we provide 4 example games and their 4 most correlated games. These correlations are very believable, and these are the games whose play probability increases the most when player mentions the game in question. The exact calculation of the probability is based on the assumption of a normal distribution, which while incorrect, seems to work well in practise.

The Tags model is based on inferring a player specific response vectors $T$ to the game tags, based on the games the player has liked. Fitting the model therefore produces a vector of $r$ interaction strengths to each of the game tags for each of the $n$ players. In Figure~\ref{fig:interpretation} we illustrate 4 example players with 3 liked games each and 4 tags with strongest implied interactions. It seems that the model is able to correctly learn the content of the liked games and describe the player in terms of their tag interactions. At prediction time, the games with these tags are predicted to be played the most by the player. The model can therefore generalize to new games by predicting games that have similar tag content.

The Questions model is based on inferring a game specific response vectors $Q$ to the player preferences, based on the players that have liked the game. Fitting the model therefore produces a vector of $s$ interaction strengths to each of the questionnaire answers for each of the $m$ games. Figure~\ref{fig:interpretation} illustrates 4 example games and 4 answers with the strongest implied interactions. The model is able to correctly learn the content of these games from the types of players that play the game, based on their questionnaire answers. At prediction time, the players with these answers are predicted to be play the game. This model can therefore generalize to new players by predicting players that have similar preferences.

The Tags X Questions model infers the interaction strengths $A$ between all player questionnaire answers and game tags, based on every player and game pair. Fitting the model produces a matrix of $r \times s$ interaction strengths, where each answer and tag pair have their own value. Figure~\ref{fig:interpretation} illustrates 4 example tags and 4 answers with the strongest implied interactions. 
These interactions are very logical and similar to what one might manually define: Puzzle tag for example interacts with the answer of liking 'Challenges of crosswords and other word puzzles'. With 61 possible answers and 379 game tags, manually defining and tuning 23 119 parameters is however infeasible. At prediction time, the players that have preferences which best match the game tags  are predicted to be play the game. This model can therefore generalize to both new players and new games simultaneously.
 
\subsection{Model Recommendations}

Finally, we illustrate the model recommendations in Table~\ref{TopRecommendations}. The MVN model produces recommendations which are close to the games the player has played, and the Tags model provides close matches in terms of game genres. The Question and Tags x Questions models rely on rather ambiguous question answers, and as a result provide quite generic recommendations. However, their answers still seem better than random. The Tags model seems better than the Questions model, even though accuracy metrics suggest the opposite conclusion.

There is a significant popularity bias visible in the Questions and Tags x Questions models, and to some extent in the MVN model. The effect of popularity can be removed from the MVN model as described earlier, and there is a similar trick that can be used with the other models. First, normalize the game like matrix $R$ column wise: substract the mean vector $\mu$ and divide by the standard deviation $\sigma=\sqrt{\mu(1-\mu)}$ to produce a matrix of standardized deviations from baseline popularity. Second, more popular games tend to have more tags provided so produce a more egalitarian 'game profile' vector by projecting $X_{\text{Tags}}$ with PCA into a lower dimensional space and normalize this vector. We found 16 dimensions worked qualitatively well. We skip reporting these results because they produced worse accuracy on the metrics, even though they virtually eliminated the effect of recommending popular but unrelated games.

\begin{table*}[htbp]
\caption{Example Players and Top 5 Game Recommendations from Different Models}
\label{TopRecommendations}
\begin{center}
\begin{tabular}{|p{8mm}|p{30mm}|p{23mm}|p{23mm}|p{23mm}|p{23mm}|p{23mm}|}
\hline
\textbf{Player} & \textbf{Questions Liked} & \textbf{Games Liked} & \textbf{MVN} & \textbf{Tags} & \textbf{Questions} & \textbf{Tags X Questions} \\
\hline
93519
&
Engaging in battle,
Weapons and skills selection for characters,
Searching and collecting rare treasures
&
Child of Light,
Dungeon Master,
Shin Megami Tensei: Persona 3
&
Persona 5,
Xenogears,
Shin Megami Tensei: Persona 4,
Chrono Cross,
Bravely Default
&
Costume Quest,
Ori and the Blind Forest,
Abyss Odyssey,
Fortune Summoners,
Bahamut Lagoon
&
World of Warcraft,
The Witcher 3: Wild Hunt,
Diablo,
The Elder Scrolls V: Skyrim,
Overwatch
&
Fallout 4,
Mass Effect 2,
Fallout 3,
Fallout: New Vegas,
Warframe
\\
\hline
93520
&
Piloting and steering vehicles,
Racing in a high speed,
Challenges of tactics
&
Project CARS,
Gran Turismo 5,
Forza Horizon 2
&
Grand Theft Auto V,
Pokémon GO,
Forza Motorsport 6,
Gran Turismo 6,
Hill Climb Racing 2
&
DiRT 4,
Gran Turismo 2,
Gran Turismo (PSP),
Forza Motorsport 4,
Forza Motorsport 2
&
Call of Duty,
Grand Theft Auto,
Clash of Clans,
Angry Birds,
Battle-field
&
StarCraft: Brood War,
StarCraft,
StarCraft II: Legacy of the Void,
Doom II RPG,
Call of Duty
\\
\hline
93521
&
Running in a fast speed while avoiding obstacles,
Developing skills and abilities,
Challenges of fast reaction
&
Shovel Knight,
Super Mario 3D World,
The Legend of Zelda: Ocarina of Time
&
The Legend of Zelda: Breath of the Wild,
Super Mario Galaxy,
Super Mario 64,
The Legend of Zelda: The Wind Waker,
The Legend of Zelda: A Link to the Past
&
The Legend of Zelda: Twilight Princess,
Rogue Legacy,
Assassin's Creed IV: Black Flag,
The Legend of Zelda: A Link to the Past,
Power Stone 2
&
TETRIS,
League of Legends,
Call of Duty,
Crash Bandicoot,
Minecraft
&
StarCraft,
Tomb Raider,
Dota 2,
StarCraft: Brood War,
Counter-Strike: Global Offensive
\\
\hline

93522
&
Hugging, kissing and making out,
Investigating the story and its mysteries,
Challenges of logical problem-solving
&
Heavy Rain,
Steins;Gate,
Life Is Strange
&
The Last of Us,
Pokémon GO,
The Witcher 3: Wild Hunt,
World of Warcraft,
TETRIS
&
Beyond: Two Souls,
The Wolf Among Us,
Zero Escape: Zero Time Dilemma,
Persona 4 Golden,
Alan Wake
&
Sudoku,
The Sims,
Angry Birds,
Pokémon GO,
Counter-Strike: Global Offensive
&
Mass Effect 2,
Fallout 3,
The Elder Scrolls IV: Oblivion,
The Elder Scrolls V: Skyrim,
Fallout: New Vegas
\\
\hline

93523
&
Decorating rooms and houses,
Hugging, kissing and making out,
Challenges of logical problem-solving
&
Cities: Skylines,
Overcooked,
The Sims 2
&
The Sims 3,
The Sims,
Civilization V,
The Sims 4,
The Elder Scrolls V: Skyrim
&
Train Valley,
Prison Architect,
The Sims,
RollerCoaster Tycoon 3: Platinum,
Tropico 4
&
Sudoku,
The Sims,
TETRIS,
Gardenscapes,
World of Warcraft
&
Warframe,
The Elder Scrolls IV: Oblivion,
Dragon's Dogma: Dark Arisen,
The Sims,
Stardew Valley
\\
\hline

\end{tabular}
\label{tab1}
\end{center}
\end{table*}

\section{Conclusion}

Research in game recommendation has focused on the prediction of missing game likes in data sets of historical player and game interactions. However, many practical tasks require predictions for completely new games and players. Collaborative filtering has been found to be a useful model in the traditional setting, but for games or players with few or no interactions different models are required. We presented content based Tags, Questions, and Tags X Questions models that generalize into new games, new players, or both simultaneously. These methods are inspired by the Singular Value Decomposition (SVD), a standard approach in collaborative filtering, where one or both of the feature vectors are assumed to be given. The optimization corresponds to a linear model with computational shortcuts, which makes them fast and easily interpretable.

We compared the following models: Random baseline, Multivariate Normal Distribution (MVN), k Nearest Neighbour (kNN) with cosine or Pearson similarity, PureSVD, SVD, Tags, Questions, Tags x Questions. We evaluated the performance with the nDCG and Precision@20 metrics within known games and players (Setting 1), new games (Setting 2), new players (Setting 3) and both new games and new players (Setting 4). We found that each content based model performed the best in the setting for which it was designed, and restricting the models to use the provided features instead of learning latent features is useful in terms of generalization ability but has a trade off in terms of accuracy. Each model can therefore be useful depending on the setting.  Finally, we note that accuracy does not tell the full story because the qualitative results do not seem to perfectly correlate with accuracy.

%




\end{document}